\documentclass{emulateapj}
\usepackage{graphicx}
\usepackage{epstopdf}
\usepackage{epsfig}

\newcommand{\hmpc}{\ensuremath{\,h^{-1}\,{\rm Mpc}\,}}
\newcommand{\ihmpc}{\ensuremath{\,h\,{\rm Mpc^{-1}}}}
     
\newcommand{\avg}[1]{\ensuremath{\langle #1 \rangle}}
\newcommand{\bma}{\begin{math}}
\newcommand{\ema}{\end{math}}
\newcommand{\beq}{\begin{equation}}
\newcommand{\eeq}{\end{equation}}
\newcommand{\beqa}{\begin{eqnarray}}
\newcommand{\eeqa}{\end{eqnarray}}
\newcommand{\bc}{\begin{center}}
\newcommand{\ec}{\end{center}} 
\newcommand{\bit}{\begin{itemize}}
\newcommand{\eit}{\end{itemize}}

\font\BFd=cmmib10
\font\BFt=cmmib10
\font\BFs=cmmib10 scaled 700
\font\BFss=cmmib10 scaled 500

\def\bbox#1{%
\relax\ifmmode
\mathchoice
{{\hbox{\BFd #1}}}
{{\hbox{\BFt #1}}}
{{\hbox{\BFs #1}}}
{{\hbox{\BFss #1}}}
\else \mbox{#1} \fi }

\def\k{{\bbox{k}}}
\def\q{{\bbox{q}}}
\def\x{{\bbox{x}}}

\def\dkc{\frac{d^3k_3}{(2 \pi)^3}}
\def\dqa{\frac{d^3q_1}{(2 \pi)^3}}
\def\dqb{\frac{d^3q_2}{(2 \pi)^3}}

\begin{document}



 
\submitted{\today. To be submitted to \apj.} 

\title{Higher Order Contributions to the 21 cm Power Spectrum}
\author{Adam Lidz\altaffilmark{1}, Oliver Zahn\altaffilmark{1,2}, Matthew McQuinn\altaffilmark{1}, Matias Zaldarriaga\altaffilmark{1,3}, Suvendra Dutta\altaffilmark{1},
Lars Hernquist\altaffilmark{1}}

\email{alidz@cfa.harvard.edu, ozahn@cfa.harvard.edu}

\altaffiltext{1}{Harvard-Smithsonian Center for Astrophysics, 60 Garden Street, Cambridge, MA 02138, USA}
\altaffiltext{2}{Institute for Theoretical Astrophysics, University of Heidelberg, Albert-Ueberle-Strasse 2, 69117 Heidelberg, Germany}
\altaffiltext{3}{Jefferson Laboratory of Physics; Harvard University; Cambridge, MA 02138, USA}

\begin{abstract}
We consider the contribution of 3rd and 4th order terms to the power spectrum 
of 21 cm brightness temperature fluctuations during the epoch of reionization (EoR), which arise 
because the 21 cm brightness temperature involves a product of the hydrogenic neutral fraction 
and the gas density. The 3rd order terms vanish for Gaussian random fields, and have been previously
neglected or ignored. We measure these higher order terms from radiative transfer simulations and 
estimate them using cosmological perturbation theory. In our simulated models, the higher order terms 
are significant: neglecting them leads to a $\gtrsim 100 \%$ error in 21 cm power spectrum
predictions on scales of $k \gtrsim 1 \ihmpc$ when the neutral fraction is 
$\avg{x_H} \sim 0.5$. At later stages of reionization, when the ionized regions are 
bigger, the higher order terms impact 21 cm power spectrum predictions on still larger scales, 
while they are less important earlier during reionization. The higher order terms have a simple physical 
interpretation. On small scales they are produced by gravitational mode coupling. Small scale structure grows more readily in large-scale overdense regions, but the same 
regions tend to be ionized and hence do not contribute to the 21 cm signal. This acts to suppress the 
influence of non-linear density fluctuations and the small-scale amplitude of the 21 cm power spectrum.
In alternate models, where the voids are reionized before over-dense regions (`outside-in' reionization), 
the effect should have the opposite sign,  and lead to an enhancement in the 21 cm power spectrum. These 
results modify earlier intuition that the 21 cm power spectrum simply traces the density power 
spectrum on scales smaller than that of a typical bubble, and imply that small scale measurements contain 
more information about the nature of the ionizing sources than previously believed. On large scales, higher order moments are not directly related to gravity. They are non-zero because over-dense regions tend to ionize first and are important in magnitude at late times owing to the large fluctuations in the neutral fraction.  Finally, we show that 
2nd order Lagrangian perturbation theory approximately reproduces the statistics of the density field from 
full numerical simulations for all redshifts and scales of interest, including the mode-coupling 
effects mentioned above. It can, therefore, be used in conjunction with semi-analytic models to accurately, 
and rapidly, explore the broad regions of parameter space relevant for future 21 cm surveys.
 \end{abstract}


\section{Introduction} \label{sec:intro}

A frontier in observational cosmology is the detection of 21 cm emission
from neutral hydrogen gas in the high redshift intergalactic medium (IGM) (e.g. Scott \& Rees 1990,
Madau et al. 1997, Zaldarriaga et al. 2004; for a review see Furlanetto et al. 2006a).
These observations promise three-dimensional information regarding the 
Epoch of Reionization (EoR), constraining the nature of the first luminous
objects and early structure formation. Indeed, several low-frequency radio
telescopes (PAST, Pen et al. 2004; MWA, http://web.haystack.mit.edu/arrays/MWA/, Bowman et al. 2006; LOFAR, http://www.lofar.org/; and SKA, http://www.skatelescope.org/), 
underway or in the planning stages, aim at detecting this signal. 

Detailed theoretical modeling is required to forecast constraints, and eventually
interpret, the results of these observations and to understand their implications
for early structure formation. The first generation of experiments will lack
the sensitivity required to make detailed 21 cm maps, and a statistical detection will
be necessary (Zaldarriaga et al. 2004, Furlanetto et al. 2004a, Morales et al. 2005, McQuinn et al. 2006). 
One statistic of choice is the power spectrum of 21 cm fluctuations, 
although other statistical measures should help in characterizing this non-Gaussian
signal (e.g. Furlanetto et al. 2004b).
It will be subtle to infer quantities like the volume-filling factor and size distribution
of HII regions from the observed 21 cm power spectrum, and more generally to extract information 
regarding the ionizing sources. 

In this paper, we consider an effect that while important for accurate calculation and interpretation of
the 21 cm power spectrum, has been neglected in many previous calculations. The 21 cm brightness 
temperature involves a product of the gas density and
the hydrogenic neutral fraction, and so the 21 cm power spectrum involves 3rd and 4th order terms. In 
Fourier space these contributions involve particular integrals of the 3 and 4-pt  functions, the bispectra and trispectra of the various fields. We will thus loosely call these terms 3 and 4-pt terms even though in real space they 
involve only two points. 

The usual intuition is that on scales much smaller than that of a typical HII region, the 21 cm
power spectrum should be proportional to the density power spectrum. This intuition ignores the
coupling between large and small scale density fluctuations which arise naturally during the non-linear
growth of structure via gravitational instability (e.g. Bernardeau et al. 2002). Indeed, mode coupling 
should give rise to non-vanishing
3-pt terms, as we detail in later sections of this paper. Qualitatively, structure grows more rapidly in
regions which are over-dense on large scales: an over-dense region acts like a closed universe with
a boosted matter density. The same regions, however, contain more sources and are ionized before 
underdense regions in our models (Sokasian et al. 2003, Furlanetto et al. 
2004a,c , Iliev et al. 2006, Zahn et al. 2006). As 
the universe reionizes, the large scale over-dense regions quickly become `dark' in a 21 cm map. 
One can think of the 21 cm field as a `masked' density field, with the ionization field playing the
role of the `mask'.
Unlike in a typical galaxy survey, however, the mask is itself correlated with the density field, 
preferentially removing large scale over-dense regions which contain boosted levels of 
small scale structure.
The upshot of this is that, in models where over-dense regions 
are ionized first, mode-couplings should suppress the contribution of density 
fluctuations to the 21 cm power spectrum. The aim of the present paper is to demonstrate this effect
and quantify its importance.

In \S \ref{sec:pk21}, we calculate the 21 cm power spectrum
(for illustrative purposes, in real as opposed to redshift space) from radiative transfer simulations,
and demonstrate the significance of 3 and 4-pt terms. In \S \ref{sec:pert} we look at the small scale effects and motivate the importance of the higher
order terms with analytic calculations based on 2nd order cosmological perturbation theory. In \S \ref{sec:large} we study the large scale limit of the higher order terms.  We then 
illustrate (\S \ref{sec:sources}) 
the dependence of the higher order terms on the properties of the ionizing sources. 
Next we (\S \ref{sec:scheme}) refine the fast numerical scheme of Zahn et al. (2005) to 
include the higher-order effects studied here.
Additionally (\S \ref{sec:evolve}), we examine how the results depend on redshift and 
ionization fraction.
Here we provide results in redshift space, generalizing the illustrative calculations of the previous 
sections.
Finally, we discuss our findings and conclude in \S \ref{sec:conclusions}.

\section{The 21 cm Power Spectrum} \label{sec:pk21}

In this section, we define terms and measure separately the contribution of low and high order terms
to the 21 cm power spectrum with radiative transfer simulations. For simplicity, we presently 
neglect the effect of peculiar velocities which we incorporate subsequently in \S \ref{sec:evolve}.
Ignoring peculiar velocities, the 21 cm brightness temperature, relative to the CMB, at observed
frequency, $\nu$, and redshift, $z$,
is (e.g. Zaldarriaga et al. 2004):
\beqa
\delta T(\nu)
\, & \approx \, & 26 \, x_H (1+\delta_\rho) \left( \frac{T_S - T_{\rm
CMB}}{T_S} \right) \left( \frac{\Omega_b h^2}{0.022} \right) \nonumber \\
& & \times \left[ \left(\frac{0.15}{\Omega_m h^2} \right) \, \left(
\frac{1+z}{10} \right) \right]^{1/2} {\rm mK}.
\label{eq:Tb}
\eeqa
In this equation, $x_H$ is the hydrogenic neutral fraction, $1 + \delta_\rho$ is the gas density in
units of the cosmic mean, $T_S$ is the spin temperature, and $T_{\rm CMB}$ is the CMB temperature.
The other symbols have their usual meanings. Throughout this work, we will make the usual simplifying
assumption that $T_S >> T_{\rm CMB}$ globally 
during reionization, implying $\delta T \propto (1 + \delta_\rho) x_H$, 
(Ciardi \& Madau 2003, Chen \& Miralda-Escud\'e 2003, Furlanetto 2006b, Pritchard \& Furlanetto 2006).

In the limit that $T_S >> T_{\rm CMB}$, and ignoring peculiar velocities, the 21 cm power spectrum
can be decomposed into the sum of several terms (generalizing the formula in Furlanetto et al. 2006c):
\begin{eqnarray}
\Delta^2_{\rm 21}(k) = \avg{T_b}^2 \avg{x_H}^2 && [\Delta^2_{\rm \delta_x, \delta_x}(k) + 2 \Delta^2_{\rm \delta_x, \delta_\rho}(k)  + 
\Delta^2_{\rm \delta_\rho, \delta_\rho}(k)\nonumber \\  + 
&& 2 \Delta^2_{\rm \delta_x \delta_\rho, \delta_x}
(k) + 2 \Delta^2_{\rm \delta_x \delta_\rho, \delta_\rho}(k) \nonumber \\ +
&& \Delta^2_{\rm \delta_x \delta_\rho, \delta_x \delta_\rho}(k)]
\label{eq:p21_decomp}
\end{eqnarray}
In this equation $\delta_x = (x_H - \avg{x_H})/\avg{x_H}$ is the fractional fluctuation in the hydrogenic 
neutral fraction, and $\avg{T_b}$ is the average 21 cm brightness temperature relative to the CMB.
Here and throughout $\Delta^2_{\rm a, b}(k)$ indicates the dimensionless
cross-power spectrum between two random fields, $a$ and $b$.  The quantity
$\Delta^2_{\rm a,a}(k) = k^3 P_{\rm a,a}(k)/(2 \pi^2)$ is
the contribution to the variance of field $a$ per $ln(k)$, and this relation defines a dimensional 
power spectrum, $P_{\rm a,a}(k)$, with our Fourier convention. The terms on the first line of Equation
(\ref{eq:p21_decomp}) are the usual low-order terms, representing the power spectrum of neutral hydrogen
fluctuations, the cross power spectrum between neutral hydrogen and gas over-density, and the
density power spectrum, respectively (e.g. Furlanetto et al. 2006c). 

The terms on the following lines, which we refer to as `higher order', are the focus of our present paper. 
Note that the term $\Delta^2_{\rm \delta_\rho, \delta_\rho}(k)$ indicates here the fully non-linear density power
spectrum. In spite of this, we will loosely refer to it, as well as the other terms on the first line of Equation
(\ref{eq:p21_decomp}) as `low order' since they contain only two fields, and 
because previous analytic 
calculations included non-linear effects for these terms using the halo model (e.g. Furlanetto et al. 2004a).
Also note that the $4$-pt term, $\Delta^2_{\rm \delta_x \delta_\rho, \delta_x \delta_\rho}(k)$, is 
non-vanishing even in the case that $\delta_\rho$ and $\delta_x$ are each
Gaussian. In this sense it is perhaps misleading to refer to it as `higher order', but we will do so
throughout for convenience. 

If we ignore the $3$-pt terms, and consider scales much smaller than that of a typical bubble, the
dominant terms from Equation (\ref{eq:p21_decomp}) are $\Delta^2_{\rm \delta_\rho, \delta_\rho}(k)$ and 
$\Delta^2_{\rm \delta_\rho \delta_x, \delta_\rho \delta_x}(k)$. In the Gaussian approximation, we will
show in the next section (Equation \ref{eq:pxrxr_2nd}) that the term 
$\Delta^2_{\rm \delta_\rho \delta_x, \delta_\rho \delta_x}(k)$ approaches 
$\Delta^2_{\rm \delta_\rho, \delta_\rho} (k) \int d\rm{lnk_3} \Delta^2_{\rm \delta_x, \delta_\x}(k_3)$ on
small scales; i.e. the expression reduces to the variance of the field $\delta_x$ multiplied by the 
density power spectrum.
Adding this $4$-pt term to the density power spectrum piece, we expect the small scale limit
of Equation (\ref{eq:p21_decomp}) to approach $\Delta^2_{\rm 21}(k) \sim \avg{T_b}^2 \avg{x_H^2} \Delta^2_{\rm \delta_\rho, \delta_\rho} (k)$. Further, to the extent that $x_H$ is either `1' or `0' at each point in
the IGM, $\avg{x_H^2} = \avg{x_H}$, implying $\Delta^2_{\rm 21}(k) \propto \avg{x_H} \Delta^2_{\delta_\rho,\delta_\rho}(k)$ on small scales. Note that if we had neglected the $4$-pt term entirely, we would have obtained a 
different small scale 
limit, $\Delta^2_{\rm 21}(k) \propto \avg{x_H}^2 \Delta^2_{\rm \delta_\rho, \delta_\rho}(k)$.

These expressions illustrate the usual intuition that, on sufficiently small scales, the 21 cm power spectrum
should trace the density power spectrum.
 In fact, we will show that the $3$-pt terms in Equation (\ref{eq:p21_decomp}) are generally substantial,
resulting in large deviations from each of these small scale limits.

\subsection{Radiative Transfer Calculations}
\label{sec:rtsims}

In order to check the effect of the `higher order' terms we 
measure each term in Equation (\ref{eq:p21_decomp}) separately, using the reionization simulations
of Zahn et al. (2006), McQuinn et al. (in prep). These simulations follow the growth of HII regions in a 
cubic box of
co-moving side length $L_{\rm box} = 65.6$ \hmpc. The Zahn et al. (2006) calculations start from an
N-body simulation run with an enhanced version of Gadget-2 (Springel 2005), tracking
$1024^{3}$ dark matter particles, and resolving dark matter halos with mass 
$M \gtrsim 2 \times 10^{9} M_\odot$. Ionizing sources are placed in simulated dark matter halos,
using a simple prescription to connect ionizing luminosity and halo mass (Zahn et al. 2006).
The simulated dark matter density field is then interpolated onto a $512^3$ Cartesian grid\footnote{This
is higher resolution than the simulations shown in Zahn et al. (2006). The higher resolution results
will be presented in McQuinn et al. (in prep).}, and we 
subsequently assume that the gas density closely tracks the simulated dark matter density field (see
Zahn et al. 2006 for a discussion).
Finally, radiative transfer is treated in a post-processing stage 
using the code of McQuinn et al. (in prep.), a refinement of the Sokasian et al. (2001, 2003) code,
which in turn uses the adaptive ray-tracing scheme of Abel \& Wandelt (2002). 

\begin{figure}
\bc
\includegraphics[width=9.2cm]{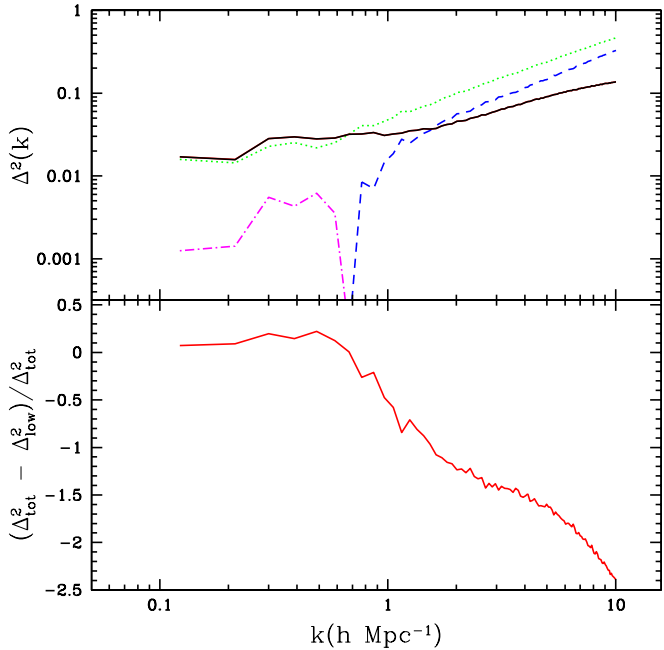}
\caption{Importance of higher-order terms for 21 cm power spectrum calculations.
{\it Top panel}: The black solid line shows the simulated 21 cm power spectrum
in real (as opposed to redshift) space at $z=6.89$, at which point the 
volume-weighted
ionization fraction is $x_{\rm i, v}=0.48$. The green dotted line shows
the contribution to the 21 cm power spectrum from the low-order terms. The magenta dot-dashed line
shows the sum of the higher order terms for wave-numbers in which the sum is positive, while
the blue dashed line shows the absolute value of the sum where it is negative. {\it Bottom panel}: The 
fractional contribution of the higher order terms as a function of scale. On small scales, our results
differ from the low-order expectation at the $\sim 100-250\%$ level.
}
\label{fig:power_decomp}
\ec
\end{figure}

The result of these calculations is shown in Figure \ref{fig:power_decomp}.
The black solid line shows the simulated 21 cm power spectrum in dimensionless units (i.e., it shows
$\Delta^2_{\rm 21}(k)/\avg{T_b}^2$ from Equation \ref{eq:p21_decomp}), the
green dotted line shows the contribution from the low order terms -- i.e., it shows the sum of
the terms on the first line of Equation (\ref{eq:p21_decomp}), and the blue dashed/magenta dot-dashed lines 
indicate the higher order terms.


The bottom panel further illustrates the importance of the 3 and 4 pt terms
for accurate predictions of the 21 cm power spectrum. The curve shows
the fractional error one makes in neglecting the higher order terms: this
error is at the $\sim 100-250\%$ level on scales of $k \sim 1-10 \ihmpc$. 
On still smaller scales the simulation results are unreliable owing to
our limited numerical resolution. As mentioned previously, we might instead
have included the Gaussian part of the $4$-pt function as a `low order' term,
in which case the low order terms amount to 
$\Delta^2_{\rm 21}(k) \propto \avg{x_H} \Delta^2_{\delta_\rho, \delta_\rho}(k)$
on small scales. Our results differ from this limit by the even larger factor
of $\sim 250\%$ at $k \sim 1 \ihmpc$. Clearly, the $3$-pt function terms are quite important in our
simulations. 

\section{Analytic Estimates} \label{sec:analytic}

\subsection{Small scales: perturbative estimates}
\label{sec:pert}

In order to develop intuition regarding the 3 and 4-pt terms, we will estimate
their form analytically, using 2nd order Eulerian perturbation 
theory (e.g. Scoccimarro 2000, Bernardeau et al. 2002). 
In this section, we restrict our analytic calculations to small scales 
($k \gtrsim 1 h$ Mpc$^{-1}$), where spatial fluctuations in the density field dominate
over fluctuations in the neutral fraction, although we will demonstrate in the next section that
the higher order terms are generally non-vanishing on larger scales as well.
In the small scale limit, we can ignore mode-couplings
between fluctuations in the hydrogenic fraction, $\delta_x$, at two different wave-numbers $\k_1$ and
$\k_2$, which should damp out on scales smaller than the typical bubble size. 

\begin{figure}
\bc
\includegraphics[width=9.2cm]{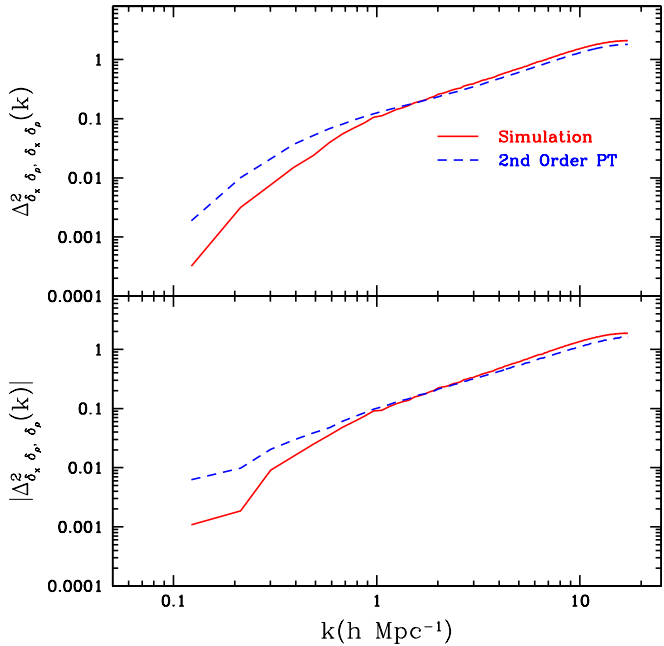}
\caption{$3$-point and $4$-point terms from our simulations, and perturbative
calculations.
{\it Top panel}: The red solid line shows the simulated $4$-pt function term. The blue dashed line
shows the $2$nd order perturbation theory calculation, according to Equation 
($\ref{eq:pxrxr_2nd}$), with
the simulated density power spectrum as input. The difference between the perturbative and simulated results 
on large scales owes
to coupling between $\delta_x$ modes, an effect ignored in our perturbative calculations. {\it Bottom panel}: A similar
comparison for the most important of the $3$-pt terms, following Equation (\ref{eq:pxrr_approx}). We show the
absolute value of this term, which is negative over all wave-numbers in our simulation. }
\label{fig:pert_theory}
\ec
\end{figure}

First, let us calculate the term $\Delta^2_{\delta_x \delta_\rho, \delta_\rho}(k)$. 
Writing
$\delta_x \delta_\rho(\k) = \psi(\k)$, and using the fact that the Fourier
transform of a product is a convolution, one has:
\begin{eqnarray}
P_{\delta_x \delta_\rho, \delta_\rho}(k) = (2 \pi)^3 \delta_D(\k_1 + \k_2) \left<\psi(\k_1) \delta_\rho(\k_2) \right> \nonumber \\
= \delta_D(\k_1 + \k_2) \int \dkc \left<\delta_x(\k_3) 
\delta_\rho(\k_1 - \k_3) \delta_\rho(\k_2)\right>
\label{eq:pxrr}
\end{eqnarray}
To lowest non-vanishing order, this expression receives contributions from expanding each
of the density field terms to 2nd order while leaving the remaining ionization field term at
1st order (see Figure \ref{fig:pert_theory} for a check on the validity of this approximation). First let us expand $\delta_\rho(\k_1 - \k_3)$ to 2nd order in
the linear density field. The 2nd order density field in Fourier space, 
$\delta^{(2)}_\rho(\k_1 - \k_3)$, is given by perturbation theory as (e.g. Scoccimarro 2000): 
\begin{eqnarray}
\delta^{(2)}_\rho(\k_1 - \k_3) = && \int \dqa \dqb \delta_D(\k_1 - \k_3 - \q_1 - \q_2) \nonumber \\
&& \times F_2(\q_1,\q_2) \delta^{(1)}_\rho(\q_1) \delta^{(1)}_\rho(\q_2)
\label{eq:2ndorder}
\end{eqnarray}

In this equation, $F_2(\q_1,\q_2)$ is the 2nd order kernel expressing mode-coupling from
non-linear evolution, $\delta^{(1)}_\rho$ denotes the 1st order density field, and 
$\delta^{(2)}_\rho$ denotes the 2nd order density field.
The second order mode-coupling kernel is given by (e.g. Scoccimarro 2000):
\begin{eqnarray}
F_2(\q_1,\q_2)= \frac{5}{7} + \frac{\q_1 \cdot \q_2}{2 q_1 q_2} \left(\frac{q_1}{q_2} + \frac{q_2}{q_1}\right) 
 + \frac{2}{7}\left(\frac{\q_1 \cdot \q_2}{q_1 q_2}\right)^2
\label{eq:f2}
\end{eqnarray}
Inserting the 2nd order expansion into the integral of Equation (\ref{eq:pxrr}),
the resulting integrand contains an expectation value of the form 
$\left<\delta_x(\k_3) \delta^{(1)}_\rho(\q_1) \delta^{(1)}_\rho(\q_2) \delta^{(1)}_\rho(\k_2)\right>$.
This 
can be rewritten as the sum of the product of two separate expectation
values, with each of two terms yielding non-vanishing contributions, 
$\left<\delta_x(\k_3) \delta^{(1)}_\rho(\q_1)\right>\left<\delta_\rho^{(1)}(\q_2) 
\delta_\rho^{(1)}(\k_2) \right>+ \left<\delta_x(\k_3) \delta^{(1)}_\rho(\q_2)\right>
\left<\delta_\rho^{(1)}(\q_1) \delta_\rho^{(1)}(\k_2) \right>$. 
The contribution to the integral of Equation (\ref{eq:pxrr}) from expanding $\delta_\rho(\k_1 - \k_3)$
to 2nd order then simplifies to $2 \delta_D(\k_1 + \k_2) P_{\rm \delta_\rho, \delta_\rho}(k_2) \int \dkc F_2(-\k_3,-\k_2) P_{\rm \delta_x, \delta_\rho}(k_3)$.
A second, similar term, arises from expanding $\delta_\rho(\k_2)$ to 2nd order. In total, our
expression for this 3-pt term to 2nd order in perturbation theory becomes
\begin{eqnarray}
P_{\rm \delta_x \delta_\rho, \delta_\rho}(k_1) =   2 \delta_D(\k_1 + \k_2) P_{\rm \delta_\rho, \delta_\rho}(k_2) \nonumber \\ \times \int \dkc F_2(-\k_3,-\k_2) P_{\rm \delta_x, \delta_\rho}(k_3)  \nonumber \\ 
+
2 \int \dkc F_2(-\k_3,\k_3-\k_1) P_{\rm \delta_x, \delta_\rho}(k_3) \nonumber \\
\times P_{\rm \delta_\rho, \delta_\rho}(\left|\k_1 - \k_3 \right|).
\label{eq:pxrr_2nd}
\end{eqnarray}

A useful further approximation results from taking the small scale limit of this equation.
The first term simplifies, after spherically-averaging to 
$(34/21) P_{\rm \delta_\rho, \delta_\rho}(k_1) \int d\rm{lnk_3} \Delta^2_{\rm \delta_x, \delta_\rho}(k_3)$.
What about the 2nd term? Since $\Delta^2_{\rm \delta_x, \delta_\rho}$ is small on small scales,
the small scale limit follows from taking $\k_1 \gg \k_3$. In this limit one can pull the density
power spectrum out of the integral. It is important to then perform the average over angles, 
before taking the $\left|\k_3\right| \rightarrow 0$ limit. This eventually yields
$(13/21) P_{\rm \delta_\rho, \delta_\rho}(k_1) \int d\rm{lnk_3} \Delta^2_{\rm \delta_x, \delta_\rho}(k_3)$.
Summing the two terms together, we arrive at a compact expression, accurate in the small-scale limit: 
\begin{eqnarray}
P_{\rm \delta_x \delta_\rho, \delta_\rho}(k_1) \sim \frac{45}{21} P_{\rm \delta_\rho, \delta_\rho}(k_1)
\int d\rm{lnk_3} \Delta^2_{\rm \delta_x, \delta_\rho}(k_3). 
\label{eq:pxrr_approx}
\end{eqnarray}

\begin{figure}
\bc
\includegraphics[width=9.2cm]{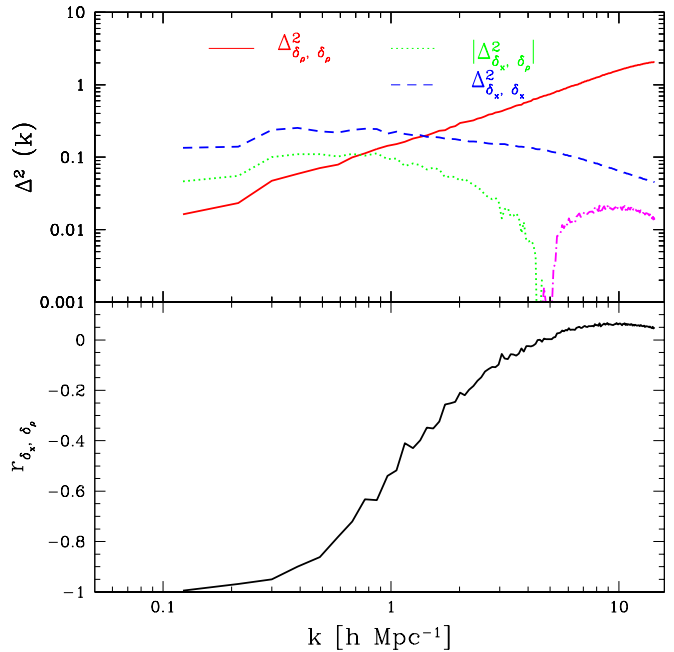}
\caption{Power spectra of density and neutral hydrogen fluctuations, and
their cross-correlation, in our simulated models.
{\it Top panel}: The red solid line shows the simulated density power
spectrum, while the blue dashed line shows the power spectrum of fluctuations
in the neutral hydrogen distribution. The green dotted line indicates the absolute
value of the cross power spectrum between neutral hydrogen fluctuations
and density fluctuations on scales where it is negative, while the magenta dot-dashed line is the same
for wave-numbers where it is positive. {\it Bottom panel}: The cross-correlation coefficient
between density fluctuations and neutral hydrogen fluctuations as a function
of wave-number. All results are at $x_{\rm i,v} = 0.48$ and $z = 6.89$.
The fluctuations in the neutral hydrogen distribution are much larger
in amplitude than the density fluctuations on large scales. The neutral
hydrogen fluctuations are perfectly anti-correlated with density 
fluctuations on large scales, while the two fields become weakly
correlated on small scales (see text).}
\label{fig:low_order}
\ec
\end{figure}

This approximate expression has an illuminating form. The integral over 
$\Delta^2_{\rm \delta_x, \delta_\rho}$ is a measure of how well the ionized regions trace
the over-densities. This integral can be written as
$\left<\delta_x \delta_\rho\right>=-1 + \left<x_H \rho\right>/\left<x_H\right> \left<\rho\right>$. Alternatively, it can be expressed in terms of the volume-weighted and mass-weighted
ionization fractions, which we denote by $x_{\rm i,v}$ and $x_{\rm i,m}$ respectively,
as $\left<\delta_x \delta_\rho\right> = (x_{\rm i,v} - x_{\rm i,m})/(1 - x_{\rm i,v})$.
This term is {\it negative} during reionization in our simulations: the ionizing sources live in over-dense
regions and reionize their surroundings, before eating out into under-dense regions. Consequently,
the mass-weighted ionization fraction always exceeds the volume-weighted ionization. 

We illustrate this explicitly in Figure \ref{fig:low_order} (see also Zahn et al. 2006), where we show 
power and cross-power spectra for density and neutral hydrogen fluctuations, as well as
the cross-correlation coefficient between neutral hydrogen and over-density, which 
is defined by $r_{\rm \delta_x, \delta_\rho}(k) = 
P_{\rm \delta_x, \delta_\rho}(k)/[{P_{\rm \delta_x, \delta_x}(k) P_{\rm \delta_\rho, \delta_\rho}(k)}]^{1/2}$. 
Here we show results from
a simulation output at $z=6.89$, at which point the simulated 
volume-weighted ionization fraction is $x_{\rm i, v}=0.48$. The figure illustrates that on large scales
the cross-correlation coefficient is always close to $-1$, a reflection of how tightly the ionized regions
track over-densities (and hence neutral hydrogen and over-density are strongly {\it anti-correlated}). 
On very small scales, the cross-correlation coefficient turns slightly positive ($r \sim 5\%$),
as ionization fronts extend further along underdense `fingers' through the 
IGM (see McQuinn et al. in prep.). Overall, however, reionization is strongly `inside-out'
in our simulations, as illustrated in the figure.
Further, note 
that $\Delta^2_{\delta_x,\delta_x}$ and $\Delta^2_{\delta_x,\delta_\rho}$
are much larger than the density power spectrum on large scales. This implies, from 
Equation (\ref{eq:pxrr_approx}), that the $3$-pt function considered here will be much larger than the 
analogous quantity for the density field: our effect is generated by $\Delta^2_{\delta_x, \delta_\rho}(k)$ on
large scales, while the density $3$-pt function is driven by the large scale density power spectrum, 
$\Delta^2_{\delta_\rho, \delta_\rho}(k)$, which is much smaller.

Indeed, at $z=6.89$,
the simulated mass-weighted ionized fraction is $x_{\rm i,m} = 0.61$, giving $\left<\delta_x \delta_\rho \right>$ 
the substantial value of $\left<\delta_x \delta_\rho\right> = -0.25$. Furthermore, the 
pre-factor in Equation 
(\ref{eq:pxrr_approx}) is $45/21$ and this term enters into the 21 cm power spectrum
(Equation \ref{eq:p21_decomp}) with an additional factor of $2$. This demonstrates that
 $2 P_{\rm \delta_x \delta_\rho, \delta_\rho}(k) \sim 
-P_{\rm \delta_\rho, \delta_\rho}(k)$ on small scales : we expect the 3-pt term to be comparable in size, and 
opposite in sign to the density power spectrum contribution. 

Strictly speaking, our perturbative expressions contain the {\it linear} density power spectrum, rather than 
the fully non-linear density power spectrum. However, we expect an expression similar to Equation (\ref{eq:pxrr_approx})
in the fully non-linear regime, containing instead the fully non-linear density power spectrum, and with a different,
perhaps scale-dependent coefficient.
Indeed, integrating Equation (\ref{eq:pxrr_2nd}) numerically with the non-linear density power spectrum as input, provides
a fairly good estimate of the small scale $3$-pt term in our simulation. We show this comparison in
the bottom panel of Figure \ref{fig:pert_theory}.\footnote{Note that the pre-factor implied
by our numerical integration is larger than the $45/21$ of Equation (\ref{eq:pxrr_approx}), 
which is accurate only in the limit of very small scales.
In other words, the 3-pt correction is even more important than implied by this approximate expression.}  
In our fiducial
model at this redshift, $\delta_x$ mode-couplings result only in a small correction to the large scale 21 cm power
spectrum (\ref{fig:power_decomp}),
although we will illustrate in \S \ref{sec:sources} and \S \ref{sec:evolve} that their effect can be significant
in other models and at different redshifts.
On small scales, the perturbative calculations provide a fairly good match to the simulation results, except
they appear to mildly underestimate the strength of the $3$-pt term at high $k$.
At any rate, our perturbative calculation clearly illustrates that the mode-coupling effect will be significant, and
demonstrates that its strength depends on how well the ionized regions trace over-densities.

Expressions for the other 3-pt term and the 4-pt term can be derived in a similar manner. The expression
for the other 3-pt term is:
\begin{eqnarray}
P_{\rm \delta_x \delta_\rho, \delta_x}(k_1) \sim \frac{34}{21} P_{\rm \delta_\rho, \delta_x}(k_1)
\int d\rm{lnk_3} \Delta^2_{\rm \delta_x, \delta_\rho}(k_3). 
\label{eq:pxrx_approx}
\end{eqnarray}
This is clearly less important than the above 3-pt term, since $P_{\rm \delta_\rho, \delta_x}(k_1) << P_{\rm \delta_\rho, \delta_\rho}(k_1)$ on small scales, although we find that this term is not completely negligible (it contributes
at the $\sim 10\%$ level on small scales).
Finally, to lowest non-vanishing order the 4-pt term is given by:
\begin{eqnarray}
P_{\rm \delta_x \delta_\rho, \delta_x \delta_\rho}&&(k_1) =   \int \dkc P_{\rm \delta_x, \delta_x}(|\k_3|) P_{\rm \delta_\rho, \delta_\rho}(|\k_1 - \k_3|)  \nonumber \\ 
&&+ \int \dkc P_{\rm \delta_x, \delta_\rho}(|\k_3|) P_{\rm \delta_x, \delta_\rho}(|\k_1 - \k_3|).
\label{eq:pxrxr_2nd}
\end{eqnarray}

The first term will dominate on small scales, and be 
roughly $P_{\rm \delta_x \delta_\rho, \delta_x \delta_\rho}(k_1) \sim P_{\rm \delta_\rho, \delta_\rho}(k_1) 
\int d\rm{lnk_3} \Delta^2_{\rm \delta_x, \delta_\x}(k_3)$.
In the top panel of Figure \ref{fig:pert_theory} we compare the simulated $4$-pt term and the results from Equation
(\ref{eq:pxrxr_2nd}),
with the simulated $P_{\rm \delta_\rho, \delta_\rho}(k)$, $P_{\rm \delta_x, \delta_x}(k)$, and 
$P_{\rm \delta_x, \delta_\rho}(k)$ as input. The agreement is comparable, although slightly worse, than for
the $3$-pt function.
At this redshift, the $4$-pt term works out to be comparable in magnitude to the dominant 
$3$-pt term, except that the $3$-pt terms come in with an additional factor of $2$ (Equation \ref{eq:p21_decomp}) 
and hence represent the dominant correction.  
Further, we have argued that this $3$-pt correction will be comparable to the contribution from the density 
power spectrum, in good agreement with the simulation results of Figure \ref{fig:power_decomp}.

\subsection{Large scales} \label{sec:large}

What about large scales? For the specific case shown in Figure \ref{fig:power_decomp} 
($z=6.89, x_{\rm i, v} = 0.48$), the higher
order terms have only a small fractional effect. However, as we detail presently and 
in \S \ref{sec:evolve},
the higher order terms are generally non-negligible even on large scales. In general, it is
challenging to calculate these terms analytically on scales where spatial fluctuations
in the neutral hydrogen distribution are comparable in strength to density fluctuations. However,
in this section we demonstrate that we can analytically estimate the higher order terms at high
ionization fractions and on large scales, when fluctuations in the neutral hydrogen 
distribution strongly dominate over density fluctuations.

\begin{figure}
\bc
\includegraphics[width=9.2cm]{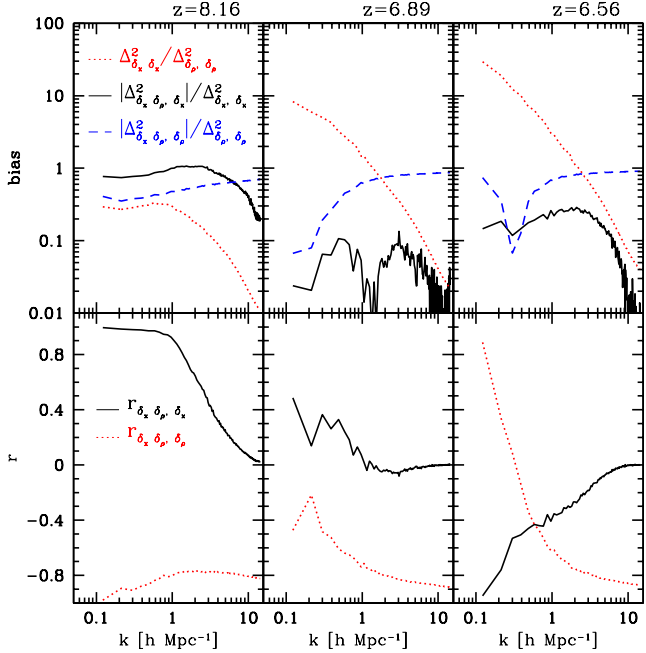}
\caption{{\it Bottom panel}: Cross correlation coefficient between
$\delta_x \delta_\rho$  with each of $\delta_x$ and $\delta_\rho$ at
(from left to right) $z=8.16, 6.89$, and $6.56$ respectively. The
volume-weighted ionized fractions at the different redshifts are
$x_{\rm i,v} = 0.11, 0.48$, and $0.70$ respectively. At early and late
times, the magnitude of the correlation coefficient is large in each
case. {\it Top panel}: Bias factors, showing the ratio of
$\Delta^2_{\delta_x,\delta_x}$ and $\Delta^2_{\delta_\rho,
\delta_\rho}$; the ratio of $\Delta^2_{\delta_x \delta_\rho,
\delta_x}$ and $\Delta^2_{\delta_x, \delta_x}$; and the ratio
$\Delta^2_{\delta_x \delta_\rho, \delta_\rho}$ over
$\Delta^2_{\delta_\rho, \delta_\rho}$. At early times (far left panel)
and on large scales, all of the various power spectra track the
density power spectrum with scale-independent bias factors. In the
intermediate redshift case (middle panel) a scale independent bias
factor appears to be a poor approximation. At late times (right
panel), $\Delta^2_{\delta_x \delta_\rho, \delta_x}$ tracks the
$\delta_x$ power spectrum closely with a roughly scale independent
bias factor.}
\label{fig:bias_all}
\ec
\end{figure}

In order to gain insight into the properties of the higher order terms on large scales, we begin
by calculating the cross correlation coefficient between $\delta_x \delta_\rho$ and
each of $\delta_\rho$ and $\delta_x$. The results of this calculation are shown in Figure 
\ref{fig:bias_all}, where we examine a range of redshifts and corresponding ionization fractions.
At $z=8.16$, when the ionization fraction is $x_{\rm i, v}=0.11$ ({\it left panel}), the figure 
illustrates that
$\delta_x \delta_\rho$ is strongly correlated with fluctuations in the hydrogenic neutral
fraction, $\delta_x$, and strongly {\em anti-correlated} with the density field, $\delta_\rho$.
Further, examining the upper leftmost panel, one can see that 
the power spectrum of neutral hydrogenic fluctuations and the $3$-pt terms each 
amount to a scale independent 
bias factor multiplied by the density power spectrum on sufficiently large scales.
At intermediate ionization fraction ($x_{\rm i,v}=0.48$, {\it middle panel}) the cross-correlation 
coefficient
between $\delta_x \delta_\rho$ and each of $\delta_x$, $\delta_\rho$ is reduced and the
$3$-pt terms are correspondingly less important, as already illustrated in Figure 
\ref{fig:power_decomp}. 

Finally, at high ionization 
fraction ($x_{\rm i, v}=0.70$, {\it right panel}), the correlation coefficients reverse sign
and are again large in magnitude. The uppermost right panel illustrates that the bias
between neutral hydrogenic fluctuations and density fluctuations, 
$\Delta^2_{\delta_x, \delta_x}/\Delta^2_{\delta_\rho, \delta_\rho}$ is large and has a strong
scale dependence. The bias between $\delta_x \delta_\rho$ and $\delta_x$, is however, relatively
independent of scale. 
This results because, at this ionization fraction, fluctuations in the hydrogenic neutral
fraction strongly dominate over density fluctuations with 
$\Delta^2_{\delta_x, \delta_x}/\Delta^2_{\delta_\rho, \delta_\rho} \sim 30$ at 
$k \sim 0.1 h$ Mpc$^{-1}$. In this case $\delta_x \delta_\rho$ just directly tracks $\delta_x$ with
a roughly scale-independent bias factor. 
It is also clear from the figure that 
$\Delta^2_{\delta_x \delta_\rho, \delta_x}$ is much greater than the other $3$-pt term,
$\Delta^2_{\delta_x \delta_\rho, \delta_\rho}$, on large scales at this ionization fraction. It is
also significantly larger than the $4$-pt term, 
$\Delta^2_{\delta_x \delta_\rho, \delta_x \delta_\rho}$, on the scales of interest, and hence
represents the dominant higher-order correction at this ionization fraction.

Can we understand analytically the weak scale dependence and strength of the bias between 
$\Delta_{\delta_x \delta_\rho, \delta_x}$ and $\Delta_{\rm \delta_x, \delta_x}$ at late times?
The following calculation will be facilitated by considering the neutral hydrogenic field
$x$, rather than examining fluctuations in the neutral hydrogenic field, $\delta_x$.
The relevant $3$-pt term, $\Delta^2_{x \delta_\rho, x}$, is related to our usual terms
by the equality:
\beqa
\Delta^2_{x \delta_\rho, x}(k) = \avg{x_H}^2 \left[\Delta^2_{\delta_x, \delta_\rho}(k) +
\Delta^2_{\delta_x \delta_\rho, \delta_x}(k) \right].
\label{eq:x_3pt}
\eeqa

We proceed to calculate $\Delta^2_{x \delta_\rho, x}(k)$. It is
simpler to understand what is going on by thinking about the
correlation function rather than the power spectrum. Thus we are
looking for a formula for the correlation $\langle x(1) \delta_\rho(1)
x(2)\rangle$ where 1 and 2 indicate two different points. We also
recall that we are working in the regime where the $x$ fluctuations
are much larger than the density ones, so correlations will be
determined by the structure of the $x$ field. For simplicity we will
consider the case when $x$ can take only the values 0 or 1. We can
then write, \beqa \langle x(1) \delta_\rho(1) x(2)\rangle && = \int
d\delta_\rho(1)  \delta_\rho(1) \times  \nonumber \\ &&
P(\delta_\rho(1),x(1)=1,x(2)=1) \nonumber \\ && = P(x(1)=1,x(2)=1)
\times \nonumber \\ && \int d\delta_\rho(1)  \delta_\rho(1)
P(\delta_\rho(1)| x(1)=1,x(2)=1). \nonumber \eeqa We now note that
P(x(1)=1,x(2)=1) is nothing other than $\langle x(1)  x(2)\rangle$. To
approximate the integral we make use of our assumption that it is the
$x$ field that dominates the correlations so that we can approximate
$P(\delta_\rho(1)| x(1)=1,x(2)=1)\approx P(\delta_\rho(1)|
x(1)=1)$. In other words we are neglecting all correlations between
the density and the neutral fraction at widely separated points other
than the ones that originate in the correlations of $x$. We thus
obtain,  \beq \langle x(1) \delta_\rho(1) x(2)\rangle  \approx \langle
x(1)  x(2)\rangle {\langle \delta_\rho  x\rangle \over \avg{x}},  \eeq
where we have used the identity: \beq \int d\delta_\rho  \delta_\rho
P(\delta_\rho| x=1)={\langle \delta_\rho  x\rangle \over \avg{x}}.  \eeq
Thus, the correlation function $\langle x(1) \delta_\rho(1)
x(2)\rangle$ has the same shape as that of $x$,  $ \langle x(1)
x(2)\rangle$.

The analytic prediction for the $3$-pt term, in the limit of strong neutral hydrogenic
fluctuations, is hence:
\beqa
\Delta^2_{x \delta_\rho, x}(k) \simeq \avg{\delta_x \delta_\rho} \Delta^2_{x, x}(k).
\label{eq:3pt_analytic}
\eeqa
A very similar argument applies to the $4$-pt function, yielding:
\beqa
\Delta^2_{x \delta_\rho, x \delta_\rho}(k) \simeq \avg{\delta_x \delta_\rho}^2 \Delta^2_{x,x}(k).
\label{eq:4pt_analytic}
\eeqa

\begin{figure}
\bc
\includegraphics[width=9.2cm]{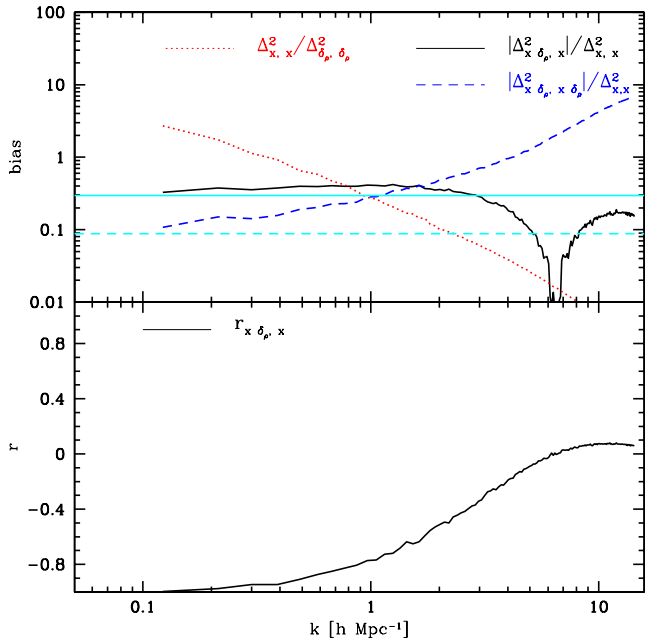}
\caption{Analytic predictions for large scale bias at $z=6.56$, $x_{\rm i, v}=0..70$. 
{\it Bottom panel}: Similar to the bottom panel
of Figure \ref{fig:bias_all}, this panel shows the cross correlation coefficient between
$x \delta_\rho$ and $x$. {\it Top panel}: The solid black line, the red dotted line, and the
blue dashed line show various bias factors, similar to the ones in Figure \ref{fig:bias_all}.
The horizontal cyan solid and dashed lines show analytic predictions for 
$\Delta^2_{x \delta_\rho, x}/\Delta^2_{x, x}$ and $\Delta^2_{x \delta_\rho, x \delta_\rho}/\Delta^2_{x, x}$ 
respectively. 
}
\label{fig:lbias_pred}
\ec
\end{figure}

In Figure \ref{fig:lbias_pred}, we compare the analytic predictions of
Equations (\ref{eq:3pt_analytic}) and (\ref{eq:4pt_analytic}) with
simulation measurements. Our simple formulae capture the effect
relatively well.

It becomes clear that in this regime the higher order terms have a
very different origin: {\it they are not related to gravity}. They
simply result from the large fluctuations in the neutral fraction and
the fact that in our models a neutral point is more likely to be underdense,
so that $\langle \delta_\rho  x\rangle$ is non-zero.

We can also obtain formulae valid in the early regime, when it is
density fluctuations that dominate the correlations. The generic
expectation value we have to calculate to compute the power spectrum
involves the probability
$P(\delta_\rho(1),x(1)=1,\delta_\rho(2),x(2)=1)$. In the early regime
we can approximate it by: \beqa P(\delta_\rho(1),&&
x(1)=1,\delta_\rho(2),x(2)=1)  \approx
P(\delta_\rho(1),\delta_\rho(2)) \times \nonumber \\ &&
P(x(1)=1|\delta_\rho(1))  P(x(2)=1|\delta_\rho(2)), \eeqa which
assumes it is the $\delta_\rho$ correlations that dominate. We can
then write: \beqa P(\delta_\rho(1),&&\delta_\rho(2)) \approx
P(\delta_\rho(1)) P(\delta_\rho(2))  \times \nonumber \\ && [1 + \xi
(r)  {\delta_\rho(1) \delta_\rho(2) \over \sigma^2} + \cdots ], \eeqa
with $\xi (r)$ the density correlation function and $\sigma^2$ its
variance.  This approximation leads to all power spectra being
proportional to that of the density. Again this is a regime were
higher order moments do not depend on gravitational non-linearities
(as we use only the lowest order form of the joint distribution
function of the density).  They are made relevant only by the large
fluctuations of $x$.

Finally, we note that the change in behavior of the correlation
coefficients seen in Figure \ref{fig:bias_all}  as reionization
proceeds can be understood as reflecting the transition between
fluctuations being dominated by $\delta_\rho$ at early times and being
dominated by $x$ at later stages. For example, Equation
(\ref{eq:x_3pt}) shows that $\Delta^2_{\delta_x \delta_\rho,
\delta_x}$ has a contribution from $\Delta^2_{\delta_x, \delta_\rho}$
and $\Delta^2_{x \delta_\rho, x}$. At early times $\Delta^2_{\delta_x,
\delta_\rho}$ dominates but becomes subdominant late during
reionization. This transition leads to the change in sign of the cross
correlation coefficients. Similar arguments apply to
$\Delta^2_{\delta_x \delta_\rho, \delta_\rho}$.

\subsection{Convergence with box-size} \label{sec:sources}

The mode-coupling effects described above imply that a rather
large simulation volume is needed to accurately simulate the small-scale
21 cm power spectrum. Indeed, owing to the significant transfer of power
from large to small scales, one might question whether even our small scale 21 cm 
power spectra results are reliable given our limited box-size
of $L = 65.6$ \hmpc. Equations (\ref{eq:pxrr_approx}), (\ref{eq:pxrx_approx}),
and (\ref{eq:pxrxr_2nd}) demonstrate that the higher-order effects depend
on two integrals: $\int d\rm{lnk_3} \Delta^2_{\delta_x, \delta_\rho}(k_3)$, and
$\int d\rm{lnk_3} \Delta^2_{\delta_x, \delta_x}(k_3)$, which represent the cross-correlation 
between $\delta_x$ and $\delta_\rho$, and the variance of $\delta_x$, respectively. 

The convergence of our small-scale 21 cm results then depend on the convergence of these two integrals with
increasing box-size. We investigate the convergence of $\avg{\delta_x^2}$ and $\avg{\delta_x \delta_\rho}$ using the 
analytic scheme of Zahn et al. (2005, 2006), which provides a quick and accurate check. 
More specifically, we apply the Zahn et al. (2005, 2006) methodology by generating a Gaussian random realization of 
a $z = 6.89$ density field in a box of
side-length $250 \hmpc$ and calculate the desired quantities. Although using a Gaussian
random density field ignores the mode-coupling effects of present interest, it does yield an accurate calculation
of $\avg{\delta_x^2}$, and $\avg{\delta_x \delta_\rho}$ (Zahn et al. 2006), which determine the convergence of 
our results with increasing
box-size. To test convergence, we compare our calculations of these quantities in the $250 \hmpc$ box
to calculations done with smaller-volume Gaussian realizations of the same density field.

\begin{figure}
\bc
\includegraphics[width=6.3cm, angle=-90]{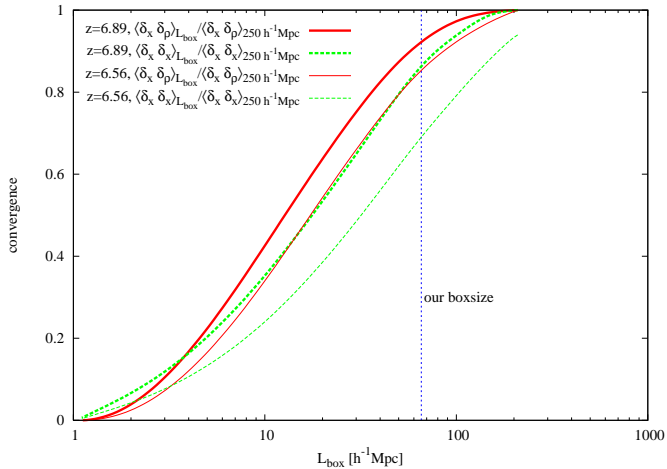}
\caption{Convergence of $\avg{\delta_x \delta_\rho}$ and $\avg{\delta_x^2}$ with increasing box-size.
These terms govern the impact of the higher order terms on small scales, and the plot hence gauges
the convergence of our simulated 21 cm power spectrum measurements on {\it small scales} with increasing box-size.
The blue dotted line indicates the size of our current simulation box, $L_{\rm box} = 65.6$ \hmpc.
The red lines show the convergence of each term for an ionization 
fraction of $x_{\rm i,v} = 0.5$, while the green lines show the convergence for $x_{\rm i,v} = 0.7$. 
With our current simulation box-size, we expect 
that $\avg{\delta_x \delta_\rho}$ is converged with an accuracy of roughly 90\%, while
$\avg{\delta_x^2}$ is converged to $\sim 80 \%$. The convergence is a bit worse when the
ionization fraction reaches $x_{\rm i,v} = 0.7$.}
\label{fig:ho_convergence}
\ec
\end{figure}

The result of this test is shown in Figure \ref{fig:ho_convergence} for ionization fractions of 
$x_{\rm i,v} \simeq 0.5$ at $z=6.89$, and $x_{\rm i, v} \simeq 0.7$ at $z=6.56$. With our 
current simulation box-size of $L_{\rm box} = 65.6 \hmpc$,
$\avg{\delta_x \delta_\rho}$ and $\avg{\delta_x^2}$ have converged to $84\%$ and $91\%$, respectively
at $x_{\rm i,v} \simeq 0.5$, which
imply that our predictions of the high $k$ 21 cm power spectrum should be relatively free of errors owing to
missing large scale power. 
For smaller box-sizes the error increases, with a box-size of $L_{\rm box} \sim 30 \hmpc$ resulting in 
a $\sim 50\%$ error. The sensitivity of $\avg{\delta_x \delta_\rho}$, $\avg{\delta_x^2}$ to large scales
arises because HII regions around individual, highly-clustered sources quickly merge into `super-bubbles' which
become quite large (Sokasian et al. 2003, Furlanetto et al. 2004a,c, Zahn et al. 2006). Our point here is that this
impacts also {\it small scale} 21 cm power spectrum predictions. Naturally, the convergence properties will
be worse at higher ionization fractions, when the HII regions are typically larger than at our fiducial
ionization fraction of $x_{\rm i, v} \sim 0.5$. This is illustrated quantitatively by
the bottom set of (green) lines in Figure \ref{fig:ho_convergence}.

\section{Dependence on Source Properties} \label{sec:sources}

The calculations in the previous section provide analytic understanding regarding the
$3$-pt functions, and demonstrate that their strength depends on 
how closely the ionization field tracks large scale over-densities.
In this section, we illustrate that this provides additional leverage in constraining
the nature of the ionizing sources and the topology of reionization from 21 cm observations. 
In particular, let us examine the higher order contribution to the 21 cm signal for three different models for 
the minimum host halo mass of the ionizing sources. 

Specifically, we consider models in
which the ionizing luminosity is proportional to the host halo mass, with sources residing in halos of mass 
larger than:  i) the cooling mass ($M_{\rm min} = 1.3 \times 10^8 M_\odot$ at $z \sim 7$, e.g. Barkana \& Loeb 2001), 
ii) $M_{\rm min} = 2 \times 10^9 M_\odot$, 
and iii) $M_{\rm min} = 4 \times 10^{10} M_\odot$, respectively. These are toy models, but they span a plausible 
range of properties given our extremely limited observational knowledge regarding the sources that 
reionized the IGM. The first model assumes
that all halos down to the cooling mass (i.e., halos with $T_{\rm vir} < 10^4 K$) contribute to reionization. The source prescriptions in ii) and iii) might, for example, approximate models in which photo-heating has limited the 
efficiency of star-formation
in small mass halos (e.g. Thoul \& Weinberg 1996, Navarro \& Steinmetz 1997, Dijkstra et al. 2004), and diminished 
their contribution to reionization. The formal halo mass resolution of our simulation is comparable to the
minimum source mass in model ii) (Zahn et al. 2006), but in model i) we use the results of McQuinn et al. (in prep.), 
which add lower mass sources into the simulation with the appropriate statistical properties. 

\begin{figure}
\bc
\includegraphics[width=9.2cm]{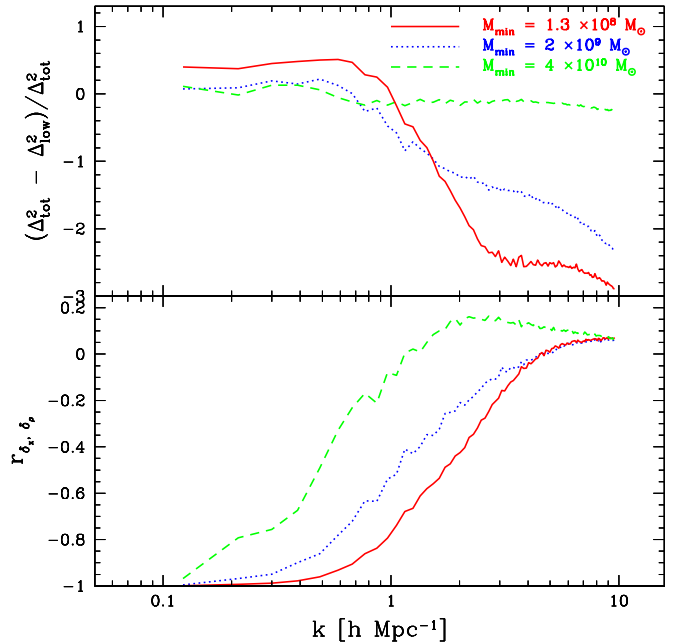}
\caption{Dependence of higher order terms on the properties of the ionizing sources.
{\it Top panel}: Similar to the bottom panel of Figure \ref{fig:power_decomp}, we
show the fractional contribution of the higher order terms as a function of scale for each
of three different source prescriptions. The effect is largest when the IGM is reionized by numerous
low mass sources.
{\it Bottom panel}: The cross-correlation between over-density and neutral hydrogen
fluctuation as a function of scale. The strength of the anti-correlation depends strongly on the source
prescription, and is largest for the cooling mass simulation.
}
\label{fig:source_prop}
\ec
\end{figure}

In Figure \ref{fig:source_prop}, we show that the impact of the higher order terms differs significantly for these
different models. In each case the source efficiency is adjusted to yield $\avg{x_H} \sim 0.5$ at 
$z \sim 7$.\footnote{More precisely our cooling mass simulation has $\avg{x_H} = 0.54$, our simulation with
$M_{\rm min} = 2 \times 10^9 M_\odot$ has $\avg{x_H} = 0.52$, and our high mass simulation has
$M_{\rm min} = 4 \times 10^{10} M_\odot$ has $\avg{x_H} = 0.53$, all at $z=6.89$.}

From the top panel it is clear that the effect depends significantly on the source properties. In the case that
the minimum mass is $M_{\rm min} = 4 \times 10^{10} M_\odot$, the effect is 
only at the $\sim 10\%$ level. 
For sources with a minimum mass of $M_{\rm min} = 2 \times 10^9 M_\odot$, as already illustrated in Figure
\ref{fig:power_decomp}, the effect is at the $\gtrsim 100\%$ level. Finally, with sources residing in halos down
to the cooling mass, the effect is even larger, at the $\gtrsim 200\%$ level.

The reason for this sensitivity is demonstrated in the bottom panel of the figure, which shows the cross-correlation
coefficient between density and neutral fraction fluctuations as a function of scale. In models where small mass halos
contribute to reionization, the ionized regions track over-densities out to smaller scales. There are two reasons for this. First, when the ionizing sources reside
in more massive halos they are more biased and produce larger bubbles at a given ionization 
fraction (Furlanetto et al. 2006c), masking out neighboring voids as well as their immediate
overdense environs. 
Next, Poisson fluctuations in the abundance of ionizing sources become increasingly important for rare, massive halos,
limiting the cross-correlation between ionization and over-density (Zahn et al. 2006). 
As a result, the 3-pt terms
(Equation \ref{eq:pxrr_approx}, \ref{eq:pxrx_approx}) become increasingly important when the ionizing sources are very abundant and more closely trace over-densities.

We emphasize that each of these models already differs in its large scale 21 cm power spectra, owing to differences in 
$\Delta^2_{\rm \delta_x, \delta_x}(k)$ and $\Delta^2_{\rm \delta_\rho, \delta_x}(k)$ between these models.
Our point here is merely that, owing to mode coupling, the small scale 21 cm power spectra gives an additional handle
on distinguishing the different models. This is clearly demonstrated by Figure \ref{fig:source_prop}.

These effects might be offset somewhat if recombinations are more important than in our simulated models. Since gas
in over-dense regions recombines faster, recombinations act to decrease the tendency for the over-dense regions to
reionize first. Further, in models where the voids reionize first, as might be the case if mini-quasars
reionize the IGM (Ricotti \& Ostriker 2004, although see Zhang et al. 2006), the
$3$-pt terms should be {\it positive} since the volume-weighted ionization fraction will exceed the mass-weighted 
ionization fraction in these models.

Finally, notice that the higher order terms contribute even on large scales, particularly in the cooling mass
model, where they amount to a $40\%$ correction (see also \S \ref{sec:evolve}). In this model, the strongest 
large scale contribution comes from the
$\Delta^2_{\rm \delta_x \delta_\rho, \delta_x}(k)$ term which is small and negative on small scales, but becomes 
significant and positive near the bubble scale (see \S \ref{sec:large} for comments on the
impact of the higher order terms on large scales).

\section{2nd order Lagrangian PT and a numerical reionization scheme.} \label{sec:scheme}
In order to forecast the ability of future 21 cm surveys to constrain reionization 
physics, we require fast and reasonably accurate predictions for the
expected 21 cm signal, spanning a wide range of model parameters. For this purpose,
rapid semi-analytic calculations are extremely valuable. One such semi-analytic scheme is that of 
Zahn et al. (2005), which is essentially a Monte-Carlo implementation of the analytic
model of Furlanetto et al. (2004a). 

The original Zahn et al. (2005) scheme uses
Gaussian random realizations of cosmological density fields, and therefore neglects
the mode-coupling effects that are the topic of our present paper. 
This semi-analytic scheme can alternatively be applied using the density field, and
halo distribution, from a full cosmological N-body simulation, as in Zahn et al. (2006). 
In this case, the approximate scheme includes the higher order effects discussed
above, and the results agree well with more detailed radiative transfer calculations.
Presently, we aim at still more rapid calculations, that additionally remove the 
expense of performing an N-body simulation. More specifically, we refine
the Gaussian random field scheme of Zahn et al. (2005) by generating cosmological density
fields according to 2nd-order Lagrangian perturbation theory (2LPT).

This ambition is plausible given that even relatively small scales
($k \lesssim 10 \ihmpc$) are still in the quasi-linear regime near
$z \sim 6$.\footnote{Although non-linear contributions to the density power spectrum
are relatively mild, higher-order contributions to the 21 cm power spectrum are 
substantial, as we demonstrated previously. The comparatively larger importance
of 3-pt terms for the 21 cm power spectrum arises because, on large scales,
$P_{\rm \delta_x, \delta_\rho}(k) \gg P_{\rm \delta_\rho, \delta_\rho}(k)$.}
Owing to this, we find that particle distributions set up using 2LPT 
accurately capture the dark matter density distribution
at the redshifts and scales of interest. 

In 2LPT, as in the Zel'dovich approximation, each particle is displaced from its 
initial Lagrangian position, $\x$, to a final Eulerian position, $\q$.  
The mapping between Lagrangian and Eulerian positions in 2LPT depends, however, on each of
the first order potential, $\phi^{(1)}(\q)$, and the 2nd order potential,
$\phi^{(2)}(\q)$. Specifically, the mapping is described by (Scoccimarro 1998):
\begin{eqnarray}
\x = \q - D_1 \nabla_\q \phi^{(1)}(\q) + D_2 \nabla_\q \phi^{(2)}(\q)
\label{eq:2nd_displace}
\end{eqnarray}
In this equation, $D_1$ and $D_2$ are the 1st and 2nd order growth
factors (Scoccimarro 1998). The particle peculiar velocities satisfy a similar 
equation (Scoccimarro 1998).
The first order potential $\phi^{(1)}(\q)$ obeys the Poisson equation
$\nabla^2 \phi^{(1)}(\q) = \delta^{(1)}(\q)$, while the 2nd order potential satisfies a separate 
Poisson equation, $\nabla^2 \phi^{(2)}(\q) = \sum_{\rm i > j} \{ \phi^{(1)}_{\rm i i}(\q) 
\phi^{(1)}_{\rm j j}(\q) - [\phi^{(1)}_{\rm i j}]^2 \}$ (Buchert et al. 1994,
Scoccimarro 1998). 

The procedure for generating particle realizations of a 2LPT density field is then straightforward.
First, one generates a Gaussian random realization of the first order density field with the appropriate
linear power spectrum. Second, one solves the first order Poisson equation for the potential,
$\phi^{(1)}(\q)$. From the first order potential, one can calculate the source term in the 2nd order
Poisson equation and subsequently solve this equation for the 2nd order 
potential, $\phi^{(2)}(\q)$. Finally,
one displaces each particle from its initial cell center using Equation (\ref{eq:2nd_displace}) with
the 1st and 2nd order potentials as input. More specifically, we generate 2LPT displacements
using  the code of Scoccimarro (1998), Crocce et al. (2006), which is now publicly 
available.\footnote{http://cosmo.nyu.edu/roman/2LPT/}

From a 2LPT particle realization, we simply interpolate to find the density field on a Cartesian grid.
We can then calculate the 21 cm power spectrum, using the ionization field
generated by applying the method of Zahn et al. (2005, 2006) to the
2LPT density field, or using the simulated ionization field.
How well do these predictions agree with results from our full N-body simulation and radiative transfer
calculation?
We can test this by generating a 2LPT realization with precisely the same first order displacements
as in our N-body simulation. 

Before presenting this comparison, we should make one caveat. The method
of Zahn et al. (2005, 2006) provides a very good, but imperfect match to the ionization field from 
more detailed radiative transfer simulations. 
Presently our aim is to test how well 2LPT
predicts the higher order terms in the 21 cm power spectrum. We thus compare the simulated and 
2LPT 21 cm fields, using in each case
the ionization field calculated from the full radiative transfer simulations. We do this comparison,
{\em rather} than using the ionization field from the radiative transfer calculation for our simulated
21 cm field, and the analytic ionization field for our 2LPT calculation. In this way, any difference
between our 2LPT and N-body calculations is attributable to differences in the density fields,
or the mode-couplings which we study presently.

\begin{figure}
\bc
\includegraphics[width=9.2cm]{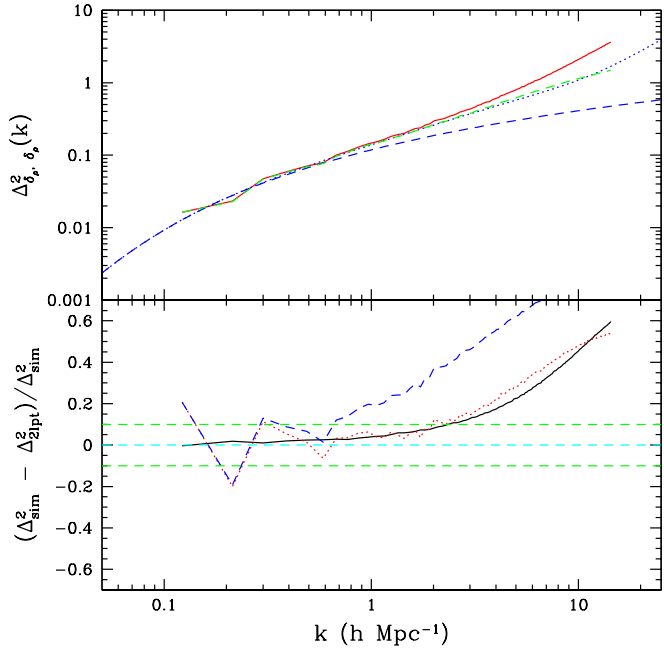}
\caption{Density power spectra from the 2LPT and N-body calculations.
{\it Top panel}: The red solid line shows the density power spectrum from our N-body simulation.
The green dashed line is the density power spectrum from the 2LPT field. The blue dashed and dotted
lines show the linear and non-linear density power spectrum according to Peacock \& Dodds (1996).
 {\it Bottom panel}: The black solid line shows the fractional difference between the simulated and
2LPT density power spectrum. The red dotted line shows the fractional difference between
the simulated power spectrum and the Peacock \& Dodds fitting formula, while the blue dashed line shows
the fractional difference between the simulated and linear theory power spectrum. The green dashed
lines indicate fractional errors of $\pm 10\%$, and the cyan dashed line runs through zero.}
\label{fig:density_power_2lpt}
\ec
\end{figure}

First we compare the simulated and 2LPT density power spectra. We use $512^3$ particles for our 2LPT particle
realization, drawn from the same initial conditions used in our N-body simulation, and 
interpolate the results onto a $512^3$ mesh using Cloud-in-Cell (CIC) interpolation. We have 
explicitly tested that our
results are relatively insensitive to the level of shell-crossing in this calculation (e.g. Scoccimarro 1998) by 
filtering our initial conditions with several different low-pass filters. For our N-body calculation we 
interpolate our $1024^3$ simulated particles onto a $512^3$ mesh.
For each of the simulated and 2LPT density
power spectra, we deconvolve the 
CIC smoothing window before calculating a binned, spherically-averaged, density 
power spectrum, and present results on scales only
where aliased high-$k$ power (e.g. Jing 2005) is insignificant. We do not attempt to subtract off shot-noise
power from our simulated results: we have run test simulations with varying particle number which 
indicate that the shot power is
significantly sub-Poisson at these redshifts (see also Springel et al. 2005), which makes it difficult to
estimate the precise level of the shot-noise power. Moreover, even in the Poisson
case, shot-power would result in a very small correction to our $1024^3$ particle simulation results : in this 
case, the
correction would be only $\sim 1\%$ at $k \sim 10 h$ Mpc$^{-1}$, and smaller on larger scales.

This comparison is shown in 
Figure \ref{fig:density_power_2lpt}, demonstrating that 2LPT provides a reasonable match to the simulated density
power spectrum, producing a $\lesssim 20\%$ underestimate of the simulated power 
at $k \lesssim 5 h$ Mpc$^{-1}$ and a $\lesssim 40\%$ underestimate
at $k \lesssim 10 h$ Mpc$^{-1}$. As a further gauge of the level of agreement, we compare the simulated density
power spectrum with the Peacock \& Dodds (1996) fitting formula. The Peacock \& Dodds (1996) power spectrum
agrees closely with the 2LPT calculation, yet appears to somewhat underestimate the simulated density power.
We emphasize that the Peacock \& Dodds (1996) fitting formula is calibrated at $z=0$, and is by no means 
guaranteed to match simulations at higher redshift. The figure demonstrates that 2LPT is, at 
any rate, just as good as this commonly used
fitting formula at $z \sim 6$ and $k \lesssim 10 h$ Mpc$^{-1}$, and is a significant improvement over a 
purely linear calculation.

\begin{figure}
\bc
\includegraphics[width=9.2cm]{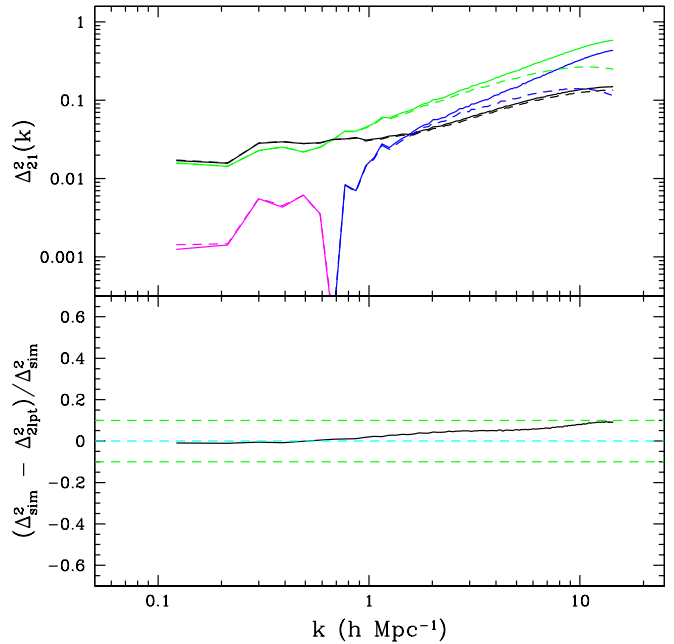}
\caption{21 cm power spectra from 2LPT and N-body calculations.
{\it Top panel (to be viewed in color)}: Similar to Figure \ref{fig:power_decomp}. The black lines show simulated
21 cm power spectra, the green lines shows the contribution to the simulated power spectra
from the low-order terms, while the blue lines show the {\it absolute value} of the sum of the 
higher-order terms. The magenta portions of these lines indicate where the sum of the higher order terms
is positive, as in Figure \ref{fig:power_decomp}. The solid lines are from the N-body calculation, while the dashed lines
show 2LPT calculations. 
 {\it Bottom panel}: The fractional difference between the simulated and N-body 21 cm fields. The
green dashed lines indicate fractional errors of $\pm 10 \%$, and the cyan dashed line runs through zero.
The agreement is generally very good.}
\label{fig:power_2lpt}
\ec
\end{figure}

Next we compare calculations of the full 21 cm power spectrum, separating out contributions from
the low-order and higher-order terms as in Equation (\ref{eq:p21_decomp}). The results of this 
calculation are shown in 
Figure \ref{fig:power_2lpt}, demonstrating that the 21 cm power spectrum calculations 
agree well (to better than $10\%$ at $k \lesssim 10 h$ Mpc$^{-1}$). The agreement is hence 
even better than the agreement found in our density power spectrum 
calculations (Figure \ref{fig:density_power_2lpt}). This superior agreement, 
however, owes somewhat to a cancellation which arises because the low-order terms in our 2LPT calculation are smaller
than in our N-body calculation, while the magnitude of the (negative) higher-order terms is also smaller in the 2LPT 
calculation, as illustrated in Figure \ref{fig:power_2lpt}. The agreement might, therefore, be a little worse at
different ionization fractions, or for different models, where this close cancellation may not occur.
On still smaller scales, where one is dominated by the $1$-halo term in the density power 
spectrum (e.g. Cooray \& Sheth 2002), 2LPT will break down further.  Additionally, our assumption that gas 
closely traces dark matter should break down (Gnedin \& Hui 1998). However, even next generation
21 cm experiments such as SKA will likely be limited to scales of $k \lesssim 10 \ihmpc$
(McQuinn et al. 2006, Bowman et al. 2006).
In conclusion, 2LPT provides a significant improvement over the
Gaussian random field scheme of Zahn et al. (2005), while requiring very little additional computational overhead.

\section{Redshift Evolution and Results in Redshift Space} \label{sec:evolve}

In the previous sections, we characterized the impact of the higher order terms, ignoring the
effect of peculiar velocities, and focusing on a single redshift for simplicity.
In this section, we expand on these calculations by incorporating the effect of peculiar 
velocities (e.g. McQuinn et al. 2006), and by 
examining the dependence of our results on ionization fraction.

In each case we calculate the spherically averaged 21 cm power spectrum; i.e. the redshift space monopole, as in
Zahn et al. (2006). In the case of the monopole, the sum of the low-order 21 cm power spectrum terms is given
by $\Delta^2_{\rm 21, low} = \avg{T_b}^2 \avg{x_H}^2 [\Delta^2_{\rm \delta_x, \delta_x} + 
(8/3) \Delta^2_{\rm \delta_x, \delta_\rho} + (28/15) \Delta^2_{\rm \delta_\rho, \delta_\rho}]$ 
(McQuinn et al. 2006, Zahn et al. 2006). In principle,
one can write down an analog to Equation (\ref{eq:p21_decomp}) in redshift space, incorporating all of the
relevant higher order terms. In practice, we instead simply calculate the full redshift space 21 cm power 
spectrum, and compare with the decomposition above.

\begin{figure}
\bc
\includegraphics[width=9.2cm, angle=0]{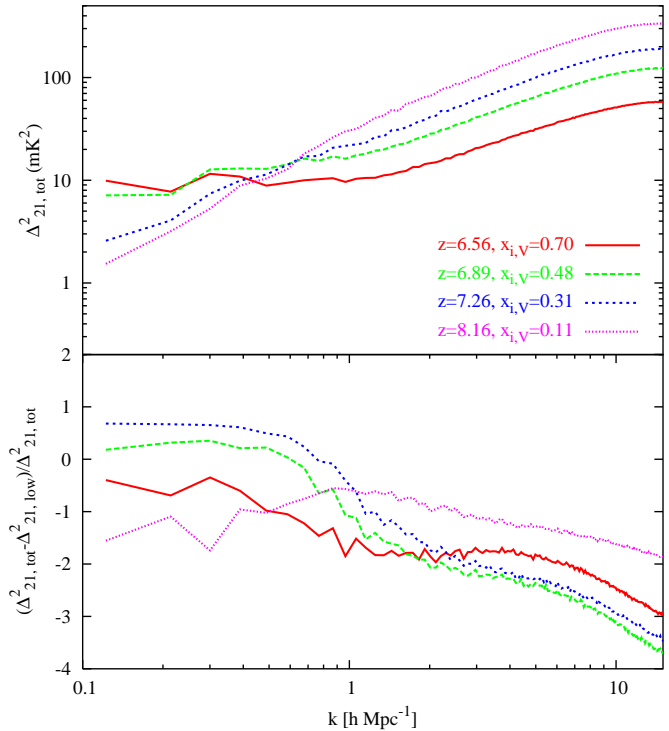}
\caption{Redshift evolution of the 21 cm power spectrum monopole in redshift space, and fractional
importance of higher order terms. {\it Top panel}: Spherically averaged 21 cm power redshift space power 
spectrum 
for different redshifts and ionization fractions. {\it Bottom panel}: Fractional importance of the higher
order terms for the spherically averaged 21 cm redshift space power spectrum.}
\label{fig:ho21_evol}
\ec
\end{figure}

The results of this calculation are shown in Figure \ref{fig:ho21_evol} for a range of different redshifts
and ionization fractions. In the top panel we show the redshift space monopole in units of $(mK)^2$, which can
be more easily compared with observational noise estimates than the dimensionless 21 cm power, which we plotted
previously for simplicity. The top panel shows the usual qualitative behavior for the 21 cm power spectrum
redshift evolution (Furlanetto et al. 2004a, Zahn et al. 2006): spatial fluctuations in the hydrogenic 
neutral fraction imprint a knee in the 21 cm power spectrum on large scales when the HII regions become sufficiently large.

The bottom panel of the Figure shows the fractional effect of the higher order terms on the 
redshift space monopole.
First, let us focus on the small scale behavior. Focusing on the $z=6.89$ curve for the moment, and comparing
with its real space analogue (Figure \ref{fig:power_decomp}), it seems that the effect is slightly enhanced in
redshift space, resulting in up to a factor of $\sim 3$ difference with the low-order calculation. 
This presumably results because peculiar velocities introduce  
additional $3$-pt terms, above and beyond the ones in Equation (\ref{eq:p21_decomp}), that are
neglected in the usual decomposition.

At high redshift, $z=8.16$, when the volume-weighted ionization fraction is
$x_{\rm i, v} = 0.11$ in our model, the small-scale suppression is less 
significant than at $z=6.89$. This results because the 
$\avg{\delta_x \delta_\rho}$ is smaller at this redshift than at our fiducial redshift 
($\avg{\delta_x \delta_\rho} = -0.12$, compared
to $\avg{\delta_x \delta_\rho} = -0.25$), amounting to a factor of $\sim 2$ less suppression 
(see Equation \ref{eq:pxrr_approx}). Finally, at the highest redshift considered here, $z=6.56$,  
where the volume-weighted
ionization fraction is $x_{\rm i, v} = 0.70$, the small-scale suppression is less significant again. This occurs
because the $4$-pt function becomes more significant at low neutral fraction, and partly compensates the $3$-pt
function suppression. Roughly speaking, the $4$-pt function term is generated by $\avg{\delta_x^2}$ while the $3$-pt
function term is generated by $\avg{\delta_x \delta_\rho}$ (see Equations \ref{eq:pxrxr_2nd} and \ref{eq:pxrr_approx}). 
Since $\avg{\delta_x^2}$ grows more rapidly with decreasing neutral fraction than $\avg{\delta_x \delta_\rho}$ 
(Zahn et al. 2006), the $4$-pt function becomes more significant relative to the $3$-pt term at 
$x_{\rm i, v} \gtrsim 0.5$, reducing the small-scale suppression, as seen in the figure.

What about the effect of the higher order terms on large scales? The
figure clearly shows that the higher order terms are generally
non-vanishing even on quite large scales, with the smallest effect
occurring at our fiducial redshift of $z = 6.89$, when the ionization
fraction is $x_{\rm i, v} \sim 0.5$. Second, note that the dependence
on ionization fraction/redshift is not monotonic, with low ionization
fractions resulting in a suppression of the 21 cm power spectrum
signal, intermediate ionization fractions yielding an enhanced signal,
and the high ionization fraction leading to a suppressed signal on
large scales again.  These large scale effects are unrelated to
gravitational instability and have their origin in the fact that in
our models high density regions tend to ionize first. Just as in the
small scale case, their amplitude is relatively large because the
fluctuations in the neutral fraction are large. The behavior of the
higher order terms on large scales depends on whether it is the density
or $x$ that dominates the correlations. The cross terms have different
signs in both limits so they go through a minimum around the mid point
of reionization (see \S \ref{sec:evolve}).

\section{Conclusions} \label{sec:conclusions}
In this paper, we demonstrated the significant impact of higher order terms on 
21 cm power spectrum predictions. While on small scales the effect originates in the mode-mode coupling induced by gravitational instability, on large scales it is unrelated to gravity.  It originates in the fact that high density regions tend to ionize first. We showed that these 
effects can help
distinguish between different models for the ionizing sources, and constrain how well
ionized regions trace large scale over-densities. Finally, we demonstrated that these
effects can be captured in semi-analytic calculations by using 2nd-order Lagrangian perturbation
theory.

How important are these effects for upcoming 21 cm surveys?
The first generation of 21 cm experiments, such as MWA and LOFAR, will likely be sensitive
only to modes with $k \lesssim 1 \ihmpc$ (Bowman et al. 2006, McQuinn et al. 2006) so only the large scale effects we discussed will be relevant. 
We showed that the higher order terms can be significant on larger scales -- to the extent
that ionized regions trace large scale overdensities -- providing important information
regarding the morphology of reionization. 
The subsequent generation of experiments, like SKA, should
extend power spectrum measurements to $k \sim 10 \ihmpc$ (Bowman et al. 2006, McQuinn et al. 2006), in which 
case our small scale effect should be extremely important, unless the IGM is reionized by very rare sources.

Another natural question is: is there a statistic that isolates the mode-coupling effect described in this paper?
As the 21 cm power spectrum is produced by several unknowns simultaneously, such a statistic would help
isolate the degree of correlation between the density and ionization fields.
We examined higher-order statistics in the vein of Zaldarriaga et al. (2001) to try and isolate our effect, but 
we find that the 21 cm $3$-pt function involves 
terms $\propto \left(P_{\rm \delta_x, \delta_x}(k)\right)_{\rm large}
\left(P_{\rm \delta_\rho, \delta_\rho}(k)\right)_{\rm small}$ -- where `large' and `small' refer to large and small scales respectively -- which are non-vanishing even in the absence of correlations between the 
density and ionization 
fields. This makes it difficult to disentangle the effect of interest, but further work in this 
direction might be interesting.

In future work, we will forecast constraints for upcoming 21 cm surveys using the 2LPT scheme presented 
here (Zahn et al. in prep). Finally, it might be interesting to investigate whether analogous $3$-pt terms are important for accurate predictions of the kinetic Sunyaev-Zel'dovich effect.

\section*{Acknowledgments} 

We thank Steve Furlanetto for helpful comments on a draft, Roman Scoccimarro for providing us with his 2LPT code and for helpful discussions, and Volker Springel
for discussions regarding the high redshift density power spectrum. 
The authors are supported by the David and Lucile Packard
Foundation, the Alfred P. Sloan Foundation, and NASA grants
AST-0506556 and NNG05GJ40G. 





\begin{thebibliography}{}

\bibitem[]{Abel02} Abel, T., \& Wandelt, B.~D. 2002, MNRAS, 330, 53

\bibitem[]{Barkana01} Barkana, R., \& Loeb, A. 2001, Phys. Rept., 349, 125

\bibitem[]{Bernardeau02} Bernardeau, F., Colombi, S., Gaztanaga, E., \& Scoccimarro, R. 2002, PhR, 367,1 

\bibitem[]{Bowman05} Bowman, J.~D., Morales, M.~F., Hewitt, J.~N. 2006, ApJ, 638, 20

\bibitem[]{Buchert94} Buchert, T., Melott, A.~L., Weiss, A.~G. 1994, A\&A, 288, 349

\bibitem[]{Chen04} Chen, X., \& Miralda-Escud\'e, J. 2004, ApJ, 602, 1

\bibitem[]{Ciardi03} Ciardi, B., \& Madau, P. 2003, ApJ, 596, 1

\bibitem[]{Cooray02} Cooray, A., \& Sheth, R. 2002, Phys. Rept., 372, 1

\bibitem[]{Crocce06} Crocce, M., Pueblas, S., Scoccimarro, R. 2006, astro-ph/0606505

\bibitem[]{Dijkstra04} Dijkstra, M., Haiman, Z., Rees, M.~J., \& Weinberg, D.~H. 2004, ApJ, 601, 666

\bibitem[]{Furlanetto04} Furlanetto, S.~R., Zaldarriaga, M., \& Hernquist, L. 2004a, ApJ, 613, 1 

\bibitem[]{Furlanetto04b} Furlanetto, S.~R., Zaldarriaga, M., \& Hernquist, L. 2004b, ApJ, 613, 16 

\bibitem[]{Furlanetto04c} Furlanetto, S.~R., Sokasian, A. \& Hernquist, L. 2004b, MNRAS, 347, 187 

\bibitem[]{Furlanetto06} Furlanetto, S.~R., Oh, S.~P., \& Briggs, F. 2006a, Phys. Rept., in press, astro-ph/0608032

\bibitem[]{Furlanetto06a} Furlanetto, S.~R. 2006b, MNRAS submitted, astro-ph/0604040

\bibitem[]{Furlanetto06b} Furlanetto, S.~R., McQuinn, M., \& Hernquist, L. 2006c, MNRAS, 365, 115

\bibitem[]{Gnedin98} Gnedin, N.~Y., \& Hui, L. 1998, MNRAS, 296, 44

\bibitem[]{Iliev06} Iliev, I.~T., Mellema, G., Pen, U.~L., Merz, H., Shapiro, P.~R., \& Alvarez, M.~A. 2006, MNRAS, 369, 1625

\bibitem[]{Jing05} Jing, Y.~P. 2005, ApJ, 620, 559

\bibitem[]{Madau97} Madau, P., Meiksin, A., \& Rees, M.~J. 1997, ApJ, 475, 429

\bibitem[]{McQuinn05} McQuinn, M., Zahn, O., Zaldarriaga, M., Hernquist, L., Furlanetto, S.~R. 2006, ApJ in press, astro-ph/0512263

\bibitem[]{Navarro97} Navarro, J.~F., \& Steinmetz, M. 1997, ApJ, 478, 13

\bibitem[]{Peacock96} Peacock, J.~A., \& Dodds, S.~J. 1996, MNRAS, 280, 19

\bibitem{Pen:2004de}
  Pen, U. L., Wu, X. P., Peterson, J., 
   arXiv:astro-ph/0404083.
  
\bibitem[]{Pritchard06} Pritchard, J.~R., \& Furlanetto, S.~R. 2006, MNRAS submitted, astro-ph/0607234

\bibitem[]{Ricotti04} Ricotti, M., \& Ostriker, J.~P. 2004, MNRAS, 352, 547

\bibitem[]{Scoccimarro98} Scoccimarro, R. 1998, MNRAS, 299, 1097 

\bibitem[]{Scoccimarro00} Scoccimarro, R. 2000, ApJ, 544, 597

\bibitem[]{Scott90} Scott, D., \& Rees 1990, MNRAS, 247, 510

\bibitem[]{sokasian01} Sokasian, A., Abel, T., \& Hernquist, L. 2001, NewA, 6,
359

\bibitem[]{sokasian03} Sokasian, A., Abel, T., Hernquist, L. \&
Springel, V. 2003, MNRAS, 344, 607

\bibitem[]{springel05} Springel, V. 2005, MNRAS, 364, 1105

\bibitem[]{springel05} Springel, V., et al. 2005, Nature, 435, 629

\bibitem[]{Thoul96} Thoul, A.~A., \& Weinberg, D.~H. 1996, ApJ, 465, 608

\bibitem[]{zahn05} Zahn, O., Zaldarriaga, M., Hernquist, L., McQuinn, M. 2005, ApJ, 630, 657

\bibitem[]{zahn06} Zahn, O., Lidz, A., McQuinn, M., Dutta, S., Hernquist, L., Zaldarriaga, M., \& Furlanetto, S.~R. 2006,
ApJ submitted, astro-ph/0604177

\bibitem[]{Zaldarriaga01} Zaldarriaga, M., Seljak, U., \& Hui, L. 2001, ApJ, 551, 48

\bibitem[]{Zaldarriaga04} Zaldarriaga, M., Furlanetto, S.~R., \& Hernquist, L. 2004, ApJ, 608, 622

\bibitem[]{Zhang06} Zhang, J., Hui, L., \& Haiman, Z. 2006, ApJ submitted, astro-ph/0607628
 
\end{thebibliography}
\end{document}